\UseRawInputEncoding
\documentclass[journal]{IEEEtran}
\usepackage[inline]{enumitem}
\usepackage{cite}
\usepackage{amsmath}
\usepackage{amsfonts}
\usepackage{multirow}
\usepackage{subfigure}
\usepackage{graphicx}
\usepackage{algorithmic}
\usepackage{array}
\usepackage{booktabs}

\begin{document}
\title{NBC2: Multichannel Speech Separation with Revised Narrow-band Conformer }

\author{Changsheng~Quan, %~\IEEEmembership{Member,~IEEE,}
        ~Xiaofei~Li%,~\IEEEmembership{Life~Fellow,~IEEE}% <-this % stops a space
\thanks{Changsheng Quan is with Zhejiang University, and Westlake University \& Westlake Institute for Advanced Study, Hangzhou, China. e-mail: quanchangsheng@westlake.edu.cn.}% <-this % stops a space
\thanks{Xiaofei Li is with Westlake University \& Westlake Institute for Advanced Study, Hangzhou, China. Corresponding author: lixiaofei@westlake.edu.cn. }% <-this % stops a space
}

% {Shell \MakeLowercase{\textit{et al.}}: Bare Demo of IEEEtran.cls for IEEE Journals}

\setlength{\doublerulesep}{0pt}

\maketitle

\begin{abstract}
This work proposes a multichannel narrow-band speech separation network.
In the short-time Fourier transform (STFT) domain, the proposed network processes each frequency independently, and all frequencies use a shared network. 
For each frequency, the network performs end-to-end speech separation, namely taking as input the STFT coefficients of microphone signals, and predicting the separated STFT coefficients of multiple speakers. 
The proposed network learns to cluster the frame-wise spatial/steering vectors that belong to different speakers. 
It is mainly composed of three components. First, a self-attention network. Clustering of spatial vectors shares a similar principle with the self-attention mechanism in the sense of computing the similarity of vectors and then aggregating similar vectors. Second, a convolutional feed-forward network. The convolutional layers are employed for signal smoothing and reverberation processing. Third, a novel hidden-layer normalization method, i.e. group batch normalization (GBN), is especially designed for the proposed narrow-band network to maintain the distribution of hidden units over frequencies. Overall, the proposed network is named NBC2, as it is a revised version of our previous NBC (narrow-band conformer) network.
Experiments show that 1) the proposed network outperforms other state-of-the-art methods by a large margin, 2) the proposed GBN improves the signal-to-distortion ratio by 3 dB, relative to other normalization methods, such as batch/layer/group normalization, 3) the proposed narrow-band network is  spectrum-agnostic, as it does not learn spectral patterns, and 4) the proposed network is indeed performing frame clustering (demonstrated by the attention maps). 
\end{abstract}

\begin{IEEEkeywords}
Multichannel speech separation, narrow-band, group batch normalization, narrow-band conformer.
\end{IEEEkeywords}

% \fi
\IEEEpeerreviewmaketitle

\section{Introduction}

% \IEEEPARstart{S}PEECH separation is to separate the speech signals of different speakers from mixture signals.
% The separated speech signals can then be used for better human hearing in scenarios like smart remote conferences, or used in other back-end tasks, such as automatic speech recognition and speaker identification, for better human-machine interactions.

% 近些年来，深度学习的方法在语音分离领域取得了巨大的进步。
% 对于单个麦克风的场景，深度学习的方法可以通过学习语音的不同的谱模式来识别不同的说话人。
% 由于单个说话人语音的稀疏性，each time frequency (TF) point of the mixture in STFT domain could be roughly considered as being dominated by a single speaker, which is called the W-disjoint orthogonality assumption \cite{yilmaz_blind_2004}.
% 假设这个假设是成立的，即忽略掉被其他说话人的能量支配的时频点，则语音可以通过估计binary mask的方式进行分离，如the well-known deep clustering \cite{hershey_deep_2016}。
% 如果我们放松这个正交性假设，使网络输出ratio masks，如phase sensistive mask ...[]，mapping或者说时域的waveform【】，则通常可以取得更好的结果。
Recently, deep learning based methods have made great progress in the field of speech separation. For scenarios with a single microphone, deep learning methods can separate different speakers by learning from the differences between spectral patterns \cite{hershey_DeepClusteringDiscriminative_2016, yu_permutation_2017}.
W-disjoint orthogonality assumption \cite{yilmaz_blind_2004} says each time-frequency (TF) bin of the mixture in short-time Fourier transform (STFT) domain could be roughly considered as being dominated by a single speaker, due to the TF sparsity of speech spectra.
Based on this assumption, speech signals can be separated by predicting a binary mask for each TF bin. The binary mask can be either directly predicted using a network, or obtained by clustering the TF embeddings as is done in the well-known deep clustering method \cite{hershey_DeepClusteringDiscriminative_2016}.
If we relax this assumption, let the network predict ratio masks \cite{yu_permutation_2017, kolbaek_MultitalkerSpeechSeparation_2017, williamson_ComplexRatioMasking_2016}, STFT coefficients \cite{lee_FullyComplexDeep_2017}, time domain waveform \cite{luo_TaSNetTimeDomainAudio_2018}, or other targets \cite{wang_SupervisedSpeechSeparation_2018}, better results can usually be obtained.
% More training targets can be found in \cite{wang_SupervisedSpeechSeparation_2018}.

% 当多个麦克风存在的时候，除了谱信息之外，神经网络还可以利用空间信息来进行分离，例如IPD...，或者学习到的空间信息【多通道unsupervised方法学习到的】，又或者多通道信号本身【时域方法、直接输入语音的方法】。
% 利用这些信息，神经网络除了可以估计前面提到的输出类型外，还可以直接预测spatial filters【fasnet, TAC】，或者分离出的多通道信号来计算spatial filter的参数【Beam-tasnet, Beam-guided tasnet, ...】。
For the multiple-microphone (multichannel) case, besides spectral information, neural networks can also leverage the spatial information of speakers, such as the widely used inter-channel phrase difference (IPD) \cite{wang_multi-channel_2018, chen_MultibandPITModel_2019}.
Besides the handcrafted spatial features (e.g. IPD), spatial information can also be explored automatically by using neural networks \cite{gu_EndtoEndMultiChannelSpeech_2019, gu_enhancing_2020} from the multichannel waveforms \cite{luo_fasnet_2019, luo_end--end_2020} or STFT coefficients \cite{wang_NeuralSpeechSeparation_2020, wang_MultimicrophoneComplexSpectral_2021}.
Besides the aforementioned targets for the single-microphone case, neural networks can also predict spatial filters directly \cite{luo_fasnet_2019, luo_end--end_2020} or predict the separated multichannel signals to estimate the spatial filters \cite{ochiai_beam-tasnet_2020, chen_BeamGuidedTasNetIterative_2022, wang_NeuralSpeechSeparation_2020, heymann_blstm_2015, wang_MultimicrophoneComplexSpectral_2021}.
% The last approach can somehow alleviate the non-linear distortion caused by the networks.

% 对于多通道语音分离，传统方法与目前的神经网络的方法存在很大的差异。
% 在生源不动的情况下（which is the assumption of most arts in the literature ）， 传统方法主要利用窄带上单个说话人的特征，如steering vector and inter-channel phase difference (IPD)，沿时间属于同一个分布来进行分离。
% 利用这一点，同时进一步假设the W-disjoint orthogonality assumption hold for simultaneous speech signals，\cite{winter_map-based_2006}提出使用层次聚类的方式对一个频带下归一化后的采样进行聚类，聚类中心即为估计出的对应频带的mixing vectors，其逆可用于unmix speech sources。
% 值得说明的是文章同时提到即使正交性假设不成立，即每个时频点也包含来自其他声源的成分，归一化后的采样点依然会在真实的mixing vectors附近形成簇。
% 类似地，方法GSS和MESSL分别利用概率模型来建模窄带上归一化后的observation，以及IPD和ILD的分布，然后通过使用EM算法来最大化观测值的likelihood，最终估计出各个说话人的mask。
% 得到的mask可以被进一步further用于计算beamforming filter的参数，如MVDR in GSS。
Beyond the deep-learning-based methods, multichannel speech separation has been intensively studied in the past several decades. One popular technique is to cluster the TF bins based on their spatial vectors (or spatial cues), as at one frequency the frames belonging to the same speaker would have identical spatial vectors (when the speaker is static). This technique is also based on the W-disjoint orthogonality assumption.
%In the case of stationary sound sources (which is the assumption of most arts in the literature), traditional methods mainly utilize the point that the features of single speaker on narrow-band, like steering vectors and IPD, along time follow the same distribution to separate. % different sources.
% With this point, clustering techniques can then be used to cluster TF bins by their underlying dominated speakers.
\cite{winter_map-based_2006} proposed to cluster the normalized TF samples (somehow equivalent to the steering vector) for each frequency independently, using a hierarchical clustering algorithm. 
% Leveraging this point and further assuming the W-disjoint orthogonality assumption \cite{yilmaz_blind_2004} hold for simultaneous speech signals, paper \cite{winter_map-based_2006} proposes to use hierarchical clustering to cluster the normalized T-F samples on a narrow-band.
%The obtained centroids are the estimated mixing vectors for that narrow-band, and then their inverse can be used to unmix speech sources.
% It should be noted that even if the W-disjoint orthogonality assumption doesn't perfectly hold, i.e. each T-F sample also has components stemmed from other sources, the normalized T-F samples can also form clusters around true mixing vectors \cite{winter_map-based_2006}.
% Similarly, besides using clustering techniques, we can estimate masks with the aforementioned point.
The model-based expectation-maximization source separation and localization (MESSL) method \cite{mandel_ModelBasedExpectation_2010} and the Guided Source Separation (GSS) method \cite{boeddecker_front-end_2018} use a probabilistic Gaussian mixture model (GMM) to model the spatial vectors (or spatial cues) of TF bins, where one Gaussian component is assigned for each speaker. The inter-channel phase/level differences (IPD/ILD) and the normalized TF samples are used in MESSL and GSS, respectively. The expectation-maximization algorithm is used to estimate the model parameters and the posterior probabilities of assigning TF bins to speakers, while the latter can be used directly for speech separation (as is done in MESSL) or for estimating the parameters of beamforming (as is done in GSS). Other multichannel speech separation techniques include beamforming \cite{gannot2001signal,gannot_consolidated_2017}, independent component analysis (ICA) \cite{makino2007blind}, etc. The foundation of these techniques is to exploit the difference of spatial correlations between different speakers. The spatial correlations are intrinsically formulated in narrow-band, in the form of steering vector, covariance matrix, IPD/ILD, etc. Correspondingly, these techniques are normally performed in narrow-band. Moreover, many other important properties are also formulated in narrow-band, for example the signal stationarity \cite{gannot2001signal} is important for discriminating between speech and noise, and the convolutive transfer function \cite{talmon2009convolutive,li2019multichannel} is widely used for modelling reverberation. 

%respectively model the normalized TF samples and inter-channel differences (IPD and ILD) on narrow-band, where one speaker in the mixture corresponds to one probabilistic model.
%Maximizing the joint likelihood of all the observations can obtain masks of each source.
% The joint likelihood of all the observations can be maximized by using Expectation-Maximization algorithm.
% Then, masks for each source can be iteratively refined during the maximization. 
%The estimated masks can further be used to calculate beamforming filters, like MVDR in GSS.
% ICA也基于这个，利用属于同一个说话人的语音成分符合同一个分布来进行分离的

In this work, we propose a narrow-band conformer network to focus on exploiting the rich information present in narrow-band, as a follow-up of our previous narrow-band LSTM (long short-term memory) networks \cite{li2019waspaa,li_narrow-band_2019,quan_MultiChannelNarrowBandDeep_2022}. 
In the STFT domain, the proposed network processes each frequency independently, and is shared by all frequencies. 
For each frequency, the network takes as input the STFT coefficients of multichannel mixture signals, and predicts the STFT coefficients of speech sources.
%It is known that one frequency has rich information for separating speech sources, such as the spectral sparsity of speech and the inter-channel differences of multiple sources.
The proposed narrow-band network is trained to learn a function/rule to automatically exploit the narrow-band information, and to perform end-to-end multichannel narrow-band speech separation. Similar to other narrow-band speech separation methods \cite{winter_map-based_2006,makino2007blind}, the proposed narrow-band method also suffers from the frequency permutation problem, namely the correspondences of separated signals at different frequencies are unclear. To solve this problem,  inspired by utterance-level permutation invariant training (uPIT) [16], we propose a full-band
PIT (fPIT) scheme that forces the separated signals of all frequencies belonging to the same speaker to locate at the same network output position. 

% 此次还需要再改进下说法。
One important functional for narrow-band speech separation is to cluster the spatial vector/feature of frames dominated by different speakers, as is done in \cite{winter_map-based_2006, mandel_ModelBasedExpectation_2010, tzinisUnsupervisedDeepClustering2019,boeddecker_front-end_2018}, under the assumption of W-disjoint orthogonality \cite{yilmaz_blind_2004}.
Clustering spatial vectors share a similar principle with the self-attention mechanism \cite{vaswani2017attention} in the sense of computing the similarity of vectors and then aggregating similar ones.
Speech signal is somehow a random process, and the estimation of its statistics, e.g. covariance matrix of multichannel speech signals, can be conducted by local smoothing/averaging operations with convolutional layers. Based on the convolutive transfer function model \cite{talmon2009convolutive,li2019multichannel}, the narrow-band microphone signal of each speaker is still a convolution between the narrow-band source signal and the convolutive transfer function, thence convolutional layers seems a natural choice to model reverberation.
% This task is simple in anechoic environment as the spatial features of speakers is not disrupted by reverberation.
% When reverberation exists, the similarity of narrow-band spatial features of single speaker will decrease, which will make the task, finding the TF components coming from the same distribution, harder.
% As a consequence, separation performance will be degraded a lot.
% But as the signals of one speaker at different time frames share the same producing parameters (room impulse response), the spatial information of a single speaker belongs to a stable random process, the statistics of which, like mean, however are less variable.
Overall, the proposed narrow-band network integrates self-attention blocks and convolutional layers, and obtains outstanding speech separation performance. The integration of self-attention blocks and convolutional layers shares a similar spirit with the Conformer network \cite{gulati2020conformer}.

% The enhanced convolution layers are expected to conduct some kind of local smoothing to find stabler spatial information which helps the attention mechanism to better find similarities.
%This integration, self-attention with enhanced convolutions, obtains extraordinary results in our experiments.
% 我们在多个数据集上进行了相关的实验，发现我们提出的分离系统的性能远超其他网络。

%%%%%%%     将这个放到方法中讲，还是在这个位置讲，感觉放到讲卷积的时候更合适    %%%%%%%%  
% 学界有很多方法实现了self-attention和卷积的组合，如 Convolutional vision Transformer (CvT) \cite{wu_cvt_2021}, Convolution-augmented Transformer (Conformer) \cite{gulati_conformer_2020}.
% 本文实现的网络与他们的主要区别在于，本文的卷积网络被置于原始Transformer的feed forward网络之内，且使用多层堆叠的方式来进一步增强卷积效果。
% 在我们之前的论文[NBC]中，我们证实了我们的卷积网络组合在窄带的框架下比Conformer取得了更好的结果。
% 在我们的预备实验中，我们也发现在我们提出的网络结构上像CvT一样将卷积置于attention的QKV计算之前，并不会带来额外的性能收益。
% 因此，我们选择了目前的组合方式。
% 更好的方式来组合注意力机制和卷积是可能存在的，但这个不是本文主要讨论的论文。
% 我们会将此留到未来的工作中进一步研究。

This work is an extension of our previously published conference paper \cite{quan_MultichannelSpeechSeparation_2022}, in which we proposed a narrow-band Conformer (NBC) network. 
The contributions of this work over \cite{quan_MultichannelSpeechSeparation_2022} include:
\begin{itemize}[leftmargin=*]
    \item We revise the narrow-band Conformer network. Specifically, two major revisions are made. First, we remove the relative positional encoding (RPE) \cite{dai_transformer-xl_2019}. In our preliminary experiments, the network with RPE trained on one speech overlap way cannot well generalize to other speech overlap ways (see Fig.~\ref{fig:ovlps} for various overlap ways), since a specific self-attention mode is formed by RPE for each overlap way. In addition, RPE needs a large amount of memory and computation resource. Second, we propose a novel normalization method for hidden units, called group batch normalization (GBN). It normalizes the hidden units of a group of training samples in one mini-batch, i.e. all frequencies belonging to the same utterance, to maintain the distribution of hidden units over frequencies. GBN can be applied in the same way for training and inference, as the frequencies of one utterance always present although they are processed independently. Our experiments show that the proposed GBN achieves a signal-to-distortion (SDR) improvement over 3 dB compared to batch normalization \cite{ioffe_BatchNormalizationAccelerating_2015}, layer normalization \cite{ba_LayerNormalization_2016} and group normalization \cite{wu_group_2018}. 
    \item The proposed method is extensively evaluated with more experiments in terms of various speech overlap ways, microphone array settings and ablation experiments. The experimental results show that the proposed method works well under various conditions, and outperforms other state-of-the-art methods by a large margin. In addition, the speaker-generalization ability is also tested, and the proposed method still works well when training with only one hour of four speakers' data.
\end{itemize}
Overall, as a revised version of NBC, the proposed network is named NBC2. Code and audio examples for the proposed method are available at \footnote{https://github.com/Audio-WestlakeU/NBSS}.

%%%%%%%%%%%%%%%%%%%%%%%% Related Works %%%%%%%%%%%%%%%%%%%%%%%%%%%%%%%
% \newpage
\section{Related Works}

\subsection{Deep Learning based Multichannel Speech Separation}
Currently, for multichannel speech separation, a large portion of advanced methods combine deep learning and beamforming techniques.
In \cite{ochiai_beam-tasnet_2020}, Beam-TasNet first estimates the multichannel speech signals for each speaker by using MC-TasNet \cite{gu_EndtoEndMultiChannelSpeech_2019}, then the minimum variance distortionless response (MVDR) beamformer is estimated for each speaker using the separated multichannel signals.
Later, Beam-Guided TasNet \cite{chen_BeamGuidedTasNetIterative_2022} added a refinement stage on Beam-TasNet to iteratively perform multichannel speech separation and MVDR beamforming.
% The drawback of MVDR beamforming is that it maintains a unit gain towards the target direction, thus even if the separated multichannel speeches are good enough, the beamformed results will inevitably contain some speech components from the interference direction.
% For that reason, VAD is applied in \cite{wang_NeuralSpeechSeparation_2020} after MVDR beamforming.
% In \cite{wang_MultimicrophoneComplexSpectral_2021} post-filtering is used after MVDR beamforming to improve the separation performance.
% 除了这类
% Instead of using the separate-then-beamforming scheme, Luo et al. proposed to predict filter-and-sum filters directly in \cite{luo_end--end_2020, luo_fasnet_2019}.
The performance of these neural beamformers are limited by the beamforming techniques, more specifically by the beam-pattern of specific beamformers. By contrast, the proposed method performs end-to-end narrow-band speech separation, thus has an unlimited performance potential, especially for the high reverberation case that beamforming techniques cannot well tackle.   

\subsection{Frequency Permutation Problem}
% Traditional multichannel speech separation methods like ICA \cite{sawada_robust_2004} and the aforementioned methods \cite{winter_map-based_2006, mandel_ModelBasedExpectation_2010, boeddecker_front-end_2018} are mainly working in narrow-band to leverage the spatial information of speakers.
%The proposed narrow-band method together with many traditional methods \cite{sawada_robust_2004, winter_map-based_2006, mandel_ModelBasedExpectation_2010, boeddecker_front-end_2018} all suffer from the well-known frequency permutation problem, i.e. how to assign the separated signals at different frequencies to the same source. 
The frequency permutation problem can be solved by leveraging the time delay of arrival (TDOA) \cite{sawada_robust_2004,mandel_ModelBasedExpectation_2010}, as the IPDs for all frequencies of the same speaker relate to one fixed TDOA. 
The inter-frequency correlation, e.g. the spectral correlation of neighbouring frequencies  \cite{mazur_approach_2009, hoffmann_UsingInformationTheoretic_2012}, can locally solve the frequency permutation problem. 
% 对于这个问题，传统上主要有估计DOA、
% To solve this problem, %Sawada et al. proposed in
% \cite{sawada_robust_2004} proposed to first use the direction of arrival (DOA) of sources to fix the permutations at some frequencies with high confidence, then use the inter-frequency correlation to decide the permutations of the remaining frequencies. 
% \cite{mazur_approach_2009} and \cite{hoffmann_UsingInformationTheoretic_2012} proposed to model each separated frequency bins with a certain distribution like generalized Gaussian, and then align the current bin with neighboring bins or the average of all corrected neighbouring bins by leveraging the similarity between their distribution parameters. 
% Mazur et al. \cite{mazur_approach_2009} proposed to model each separated frequency bins using the generalized Gaussian distribution (GGD) and align neighboring bins by leveraging the tiny differences of the parameters of the GGD between neighboring bins of the same speaker.
% Similarly, Hoffmann et al. \cite{hoffmann_UsingInformationTheoretic_2012} tried Rayleigh distribution to model the magnitudes of spectral components and GGD to model the logarithms of the speech magnitudes.
% When to align one frequency, they don't use the statistics of only one of its adjacent frequency as a middle misalignment might cause the permutations totally wrong, instead they use the statistics of the average of all corrected neighbouring frequencies.
In \cite{ito_PermutationfreeClusteringRelative_2015}, Ito et al. proposed a permutation-free clustering method for blind source separation, using the common amplitude modulation property of speech, i.e. the frequency components of a speech signal tend to be synchronously activated, to bind the source dominance priors of different frequencies of the same speaker.
% set the source dominance prior as time-variant but frequency-invariant so that permutation-free is achieved through the binding (frequency-invariant) of the source dominance priors of different frequencies of the same speaker.
% 最近，Nobutaka等人提出了一个p-free的方法。为了达到p-free，他们使用了语音的common amplitude modulation特性，即the frequency components of a speech signal tend to be synchronously activated。利用这个特性，他们设置source dominance prior为time-variant but frequency-invariant. 如此，通过绑定不同频率的source dominance prior，他们实现了p-free。
The proposed full-band PIT is similar to this permutation-free method \cite{ito_PermutationfreeClusteringRelative_2015}, in the sense that the predictions of one speaker is also bound across frequencies. To resolve the permutation, the network may partially leverage the common amplitude modulation property, and possibly the TDOA as well.
%The difference is that we don't use the common amplitude modulation property of speech, instead we just bind the predictions at the same output stream as predictions of one speaker, and let the network to automatically find the best permutation solution from the perspective of speech quality.
% This task for the neural network is similar to the task of narrow-band localization, but simpler to the localization task as the network only needs to decide the output channels with the directional information provided by the narrow-band mixtures, doesn't need to give the precise direction.
% 本文提出的解决p问题的方法和这个方法类似，但是我们没有使用语音的common amplitude modulation特性，而是在预测值上进行频率绑定，使用网络从最终信号质量的角度，来自动寻找解决FPP问题的最优解。神经网络在训练时，其行为非常类似与在窄带上进行定位，但是比定位更简单，因为其只需要根据方位信息来决定该窄带上各个预测值的输出通道，而不需要给出方位的具体值。

% \subsection{Permutation Invariant Training}
% % 介绍一下目前的PIT的进展。说一下Full-band PIT与其他PIT的区别【和FPP问题相关】

% \subsection{Integration of Convolution and Transformer}
% 题目改成：conformer？
% 卷积对于语音是比较重要的一个设计，能够进行局部的平滑，得到更稳定的特征，如本文中就期望使用卷积来得到更稳定的空间信息的表示。
% 主要说明和conformer的区别

% 区别就包括EN

\subsection{Network Normalization}
% Hidden units normalization is of great importance for the training convergence speed of neural networks.
% 网络中的hidden的归一化是对于网络的训练、收敛是非常重要。
% Without normalization, the parameter changes in low layers will cause a large variation to the input distribution of high layers, and the higher the larger.
% This phenomenon is the so called internal covariate shift \cite{ioffe_BatchNormalizationAccelerating_2015}.
% 如果在深层网络中不使用归一化，则使用梯度更新参数之后，浅层网络参数的变化会导致深层网络的输入分布发生巨大变化，也即是所谓的internal covariate shift.

Batch normalization (BN) \cite{ioffe_BatchNormalizationAccelerating_2015}, layer normalization (LN) \cite{ba_LayerNormalization_2016} and group normalization (GN) \cite{wu_group_2018} are three main widely used normalization methods.
BN is proposed to solve the internal covariate shift problem, i.e. the parameter changes in low layers may cause large variations to the input distribution of high layers.
During training, BN uses the samples in one mini-batch to calculate the statistics used for  normalization, which are regarded as approximations of the statistics of the whole population.
At test time, the population statistics are also used, but are approximated by the moving average of the statistics obtained in the training process. BN usually requires a large training batch size to have a good statistical approximation \cite{wu_group_2018}. In addition, BN is not suitable for recurrent neural networks as it  requires different statistics for different time steps, but the the number of time steps varies from sequence to sequence \cite{ba_LayerNormalization_2016}. 
%Therefore, BN cannot be applied to extremely large distributed models where the mini-batches have to be small \cite{ba_LayerNormalization_2016}.
LN and GN are proposed to mitigate the problems of BN.
The statistics in LN are calculated over all the hidden units of one layer of one training sample.
While in GN, the hidden units of one layer are divided into groups, and the statistics are calculated over the hidden units for each group. The hidden units in one group are more correlated to each other than to the hidden units of other groups, and the group-wise normalization is benefit for maintaining their correlations. Different from BN, LN and GN are irrelevant to batch size, and can use a small batch size.

The proposed group batch normalization (GBN) targets a situation that is not considered in BN, LN and GN, namely a group of highly correlated data, e.g. the different frequencies of one utterance in this work, are independently processed by the network and treated as independent training samples. This group of data always present together no matter for training or inference. GBN normalizes over this group of data to maintain their correlations. The situation that a group of highly correlated data are independently processed happens in many other frameworks as well, such as the intra-layers in the dual-path framework \cite{luo_DualPathRNNEfficient_2020, dang_DPTFSNetDualPathTransformer_2022, subakan_attention_2021}, or the subband layers in FullSubNet \cite{hao_FullsubnetFullBandSubBand_2021}, for which the proposed GBN strategy may also be applicable. 

\section{Multichannel Narrow-band Speech Separation}

\begin{figure*}[htbp]
  \centering
  \includegraphics[width=\linewidth]{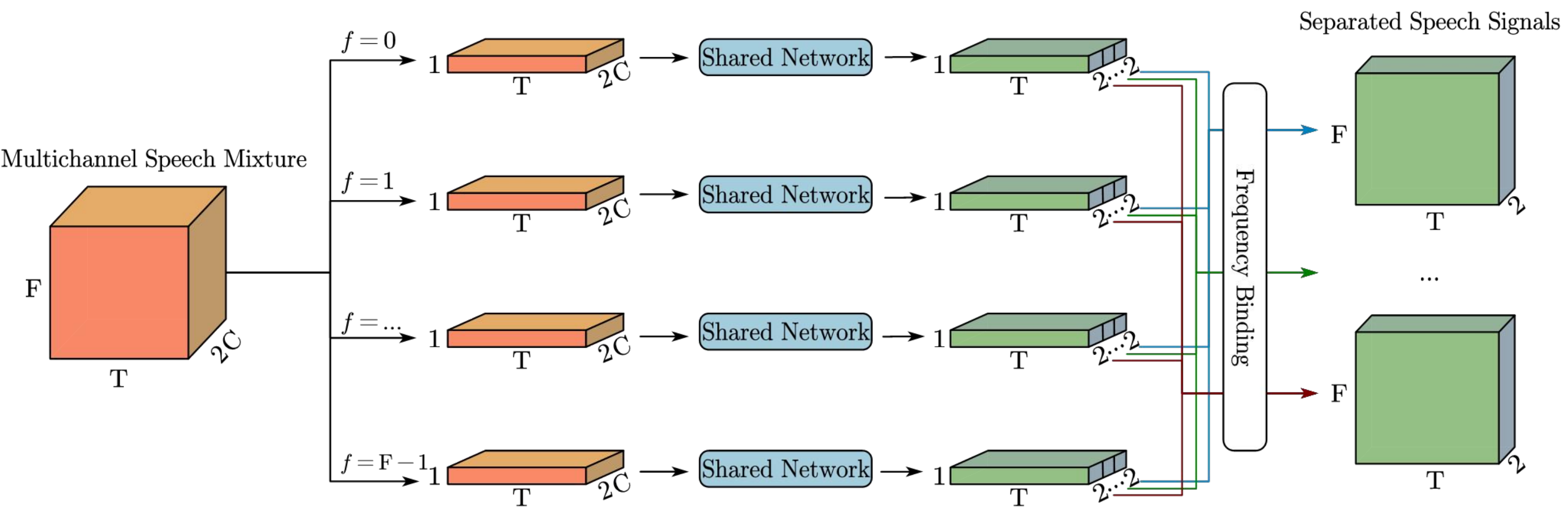}
  \caption{Framework of narrow-band deep speech separation. Frequencies are processed independently using a shared network. Then the separated results at the same output position of different frequencies are bound together to form full-band spectra of each speaker.}
  \label{fig:flowchart}
\end{figure*}

This section introduces the proposed multichannel narrow-band speech separation framework. Specifically, we consider multichannel signals in the STFT domain: 
\begin{equation}
  {\rm X}_{f,t}^{c} = \sum_{n=1}^N {\rm Y}_{f,t}^{n,c},
  \label{eq1}
\end{equation}
where ${\rm X}$ and ${\rm Y}$ are the complex-valued STFT coefficients of microphone signals and of the reverberant spatial image of speech sources, respectively.
$f\in \{0,...,F-1\}$, $t\in \{1, ..., T\}$, $c$ $\in$ $\{1,...,C\}$, and $n\in \{1,...,N\}$ denote the indices of frequency, time frame, microphone channel and speaker, respectively.
This work aims to recover the reverberant spatial image of each speaker at a given reference channel, e.g.\ ${\rm Y}_{f,t}^{n,r}$ with $r$ denoting the reference channel.
% For simplicity, the index of reference channel of spatial images is omitted in the following text.

The proposed narrow-band deep speech separation framework is shown in Fig. \ref{fig:flowchart}. 
It first separates the speech mixture for each frequency independently by using a shared network, then the separated results of all frequencies are bound together to form the full-band spectra of each speaker for solving the frequency permutation problem.

\subsection{Narrow-band Deep Speech Separation}
\label{sec:nbss}

As shown in Fig. \ref{fig:flowchart}, speech separation is performed independently for each frequency, and the same separation network (will be presented in Section \ref{sec:net_arch}) is shared by all frequencies.

The network takes the STFT coefficients of a single frequency as its input sequence:
\begin{equation}
  \mathbf{X}_{f} = (\mathbf{X}_{f,1},\dots,\mathbf{X}_{f,T}) \in \mathbb{C}^{C\times T},
\end{equation}
where $\mathbf{X}_{f,t} = [{\rm X}_{f,t}^1,\dots,{\rm X}_{f,t}^C]^{T} \in \mathbb{C}^{C\times 1}$
denotes the concatenation of the multichannel STFT coefficients of one TF bin.
The output of the network is the sequence of separated speech signals for the same frequency, denoted by $\boldsymbol{\widehat{{\rm Y}}}_{f}\in\mathbb{C}^{N\times T}$, which is the prediction of the ground truth signal of $N$ speakers $\boldsymbol{{{\rm Y}}}_{f}\in\mathbb{C}^{N\times T}$.
Here we denote the input and output sequence in complex domain for notational simplicity, while their real and imaginary parts are actually used in real implementation, with vector dimensions of $2C$ and $2N$, respectively.

Magnitude normalization is performed for each frequency to facilitate the training of the network, as $\mathbf{X}_{f}/ \overline{{\rm X}}_f$, where $\overline{{\rm X}}_{f} = \frac{1}{T}\textstyle\sum_{t=1}^T |{\rm X}_{f,t}^r|$.
An inverse normalization is applied to the network output as $\boldsymbol{\widehat{{\rm Y}}}_{f} \overline{{\rm X}}_{f}$ to recover the original magnitude of each frequency.

\subsection{Full-band Permutation Invariant Training}
End-to-end training of the narrow-band network needs to solve both the label permutation problem and the frequency permutation problem.
The label permutation problem can be directly solved at the frequency level by applying the widely used PIT technique \cite{yu_permutation_2017, kolbaek_MultitalkerSpeechSeparation_2017} for each frequency.
But applying PIT for each frequency separately, although the speech signals can be well separated at each frequency, it still suffers from the frequency permutation problem, as is for traditional narrow-band methods \cite{winter_map-based_2006, gannot_consolidated_2017, boeddecker_front-end_2018}.

% The frequency permutation problem is to find out which F predictions at different frequencies can form the complete spectra of the same speaker, which should be solved for end-to-end training and testing.

To solve the frequency permutation problem and label permutation problem together, we propose a frequency binding technique, which forces the network to produce predictions with identical speaker label permutation for all frequencies.
Specifically, the predictions at the same output position of all frequencies, 
i.e.\ the prediction of all frequencies at the $n$-th output position $\widehat{\mathbf{Y}}^n = [\widehat{\mathbf{Y}}_0^n;\dots;\widehat{\mathbf{Y}}_{F-1}^n]\in \mathbb{C}^{F\times T}$,
are forced to belong to the same speaker, and bound together for the calculation of the PIT loss for solving the label permutation problem.
We call this integration of frequency binding and permutation invariant training as full-band PIT (fPIT) as it calculates the loss using all the frequency bands.
The fPIT is then defined as: 

\begin{small}
\begin{equation}
  \text{fPIT}(\boldsymbol{\rm \widehat{Y}}^{1},\ldots, \boldsymbol{\rm \widehat{Y}}^{N}, \boldsymbol{\rm {Y}}^{1},\ldots,\boldsymbol{\rm {Y}}^{N}) \\
  =
  \mathop{min}_{p\in \mathcal{P}}
  \frac{1}{N}
  \sum_n \mathcal{L}(
    \boldsymbol{{\rm {Y}}}^n
    ,
    \boldsymbol{{\rm \widehat{Y}}}^{p(n)}
  ) 
  \label{eq7}
\end{equation} 
\end{small}

\noindent where $\boldsymbol{{\rm {Y}}}^n \in \mathbb{C}^{F\times T}$ denotes the ground truth STFT coefficients of the $n$-th speaker at the reference channel.
$\mathcal{P}$ denotes the set of all possible permutations, and $p$ denotes a permutation in $\mathcal{P}$ which maps the ground truth labels to the prediction labels.
$\mathcal{L}$ denotes a loss function.

% 使用frequency binding解决fpp问题是合理的。
% 其解决的问题类似于在窄带上做定位，但是比窄带定位更加简单，因为需要分辨的方位只有有限的N个，且不需要输出具体的方位值。
% The frequency binding in fPIT requires the network to not only separate the speech signals for each individual frequency,
% but also output the predictions of all frequencies for one speaker at the same position,
% even though the network processes frequencies separately.
% The former task relies on learning narrow-band spatial information to discriminate multiple speakers.
% The latter task possibly uses the (partial) narrow-band spatial information that are consistent along frequencies to determine the output position, such as the inter-channel cues related to the speaker direction.

For training, we use the negative of SI-SDR \cite{roux_sdr_2019} as the loss function of the proposed method:
% \begin{small}
\begin{equation}
%  \begin{aligned}
   \mathcal{L}(\mathbf{Y}^n,\mathbf{\widehat{\mathbf{Y}}}^{p(n)}) 
%   &= - \text{SI-SDR}(\boldsymbol{{\rm y}}^n,\boldsymbol{{\rm \widehat{y}}}^{p(n)}) \\
   = -10 \log_{10}\frac{\left \| \alpha \boldsymbol{{\rm y}}^n \right \|^2}{\left \| \alpha \boldsymbol{{\rm y}}^n - \boldsymbol{{\rm \widehat{y}}}^{p(n)} \right \|^2}
  \label{eq8}
% \end{aligned}
\end{equation}
% \end{small}
where $\alpha=({\boldsymbol{{\rm \widehat{y}}}^{p(n)}})^T\boldsymbol{{\rm y}}^n/\left \| \boldsymbol{\rm y}^n \right \|^2$, $\boldsymbol{{\rm y}}^n$
and $\boldsymbol{{\rm \widehat{y}}}^{p(n)}$ are the inverse STFT of $\boldsymbol{{\rm Y}}^n$ and $\boldsymbol{{\rm \widehat{Y}}}^{p(n)}$, respectively.

\section{Network Architecture of Narrow-band Conformer}
\label{sec:net_arch}

Consider the speech signal of one speaker in the STFT domain, using the narrow-band approximation \cite{gannot_consolidated_2017}, we have $\textbf{Y}^n_{f,t} \approx S^n_{f,t}\textbf{A}^n_f$, where $S^n_{f,t}$ and $\textbf{Y}^n_{f,t}\in \mathbb{C}^{C\times 1}$ respectively denote the speech signal of the $n$-th speaker and its multichannel spatial images at frequency $f$ and frame $t$, and $\textbf{A}^n_f \in \mathbb{C}^{C\times 1}$ is the acoustic transfer function from the $n$-th speaker to microphones at frequency $f$. Note that, this work only considers the static speaker case, for which $\textbf{A}^n_f$ is time independent. Based on the W-disjoint orthogonality assumption \cite{yilmaz_blind_2004}, namely each T-F bin is dominated by one speaker, one effective way \cite{winter_map-based_2006} to perform speech separation is to cluster the frames using the acoustic transfer function (or spatial/steering vector) estimated at each frame, as different speakers have different spatial vectors. 
%With this approximation, we can see that the single frequency speech components of one speaker at different frames share the same acoustic transfer function, thus the inter-channel spatial clues of one speaker like IPD are similar along time in a single frequency.
%Relying on the similarity of inter-channel spatial clues and the spectral sparsity of speech, narrow-band speech separation can be performed by spatial vector clustering to group frames dominated by the same speaker, as is done in \cite{winter_map-based_2006}.
From the perspective of computing the similarity of vectors, spatial vector clustering shares a similar principle with self-attention mechanism, which motivates us to employ self-attention in our proposed network.

Speech signal is somehow a random process, and the estimation of spatial vector relies on the computation of speech statistics, such as the covariance matrix of multichannel speech signals. This motivates us to use convolution layers to perform local smoothing/averaging operations for the computation of speech statistics. In addition, the narrow-band approximation holds only when RIR is short relative to the STFT window, which is usually not the case when reverberation is large. As a more precise model, such as the convolutive transfer function approximation \cite{talmon2009convolutive,li2019multichannel}, $\textbf{Y}^n_{f,t}$ is still a convolution between $S^n_{f,t}$ and the STFT-domain representation of RIR. Although this work does not conduct dereverberation, the capability of modelling reverberation is still important for separating reverberant speech. It seems a natural choice to use convolutional layers for modelling reverberation.

\begin{figure}[t]
  \centering
  \includegraphics[width=0.8\linewidth]{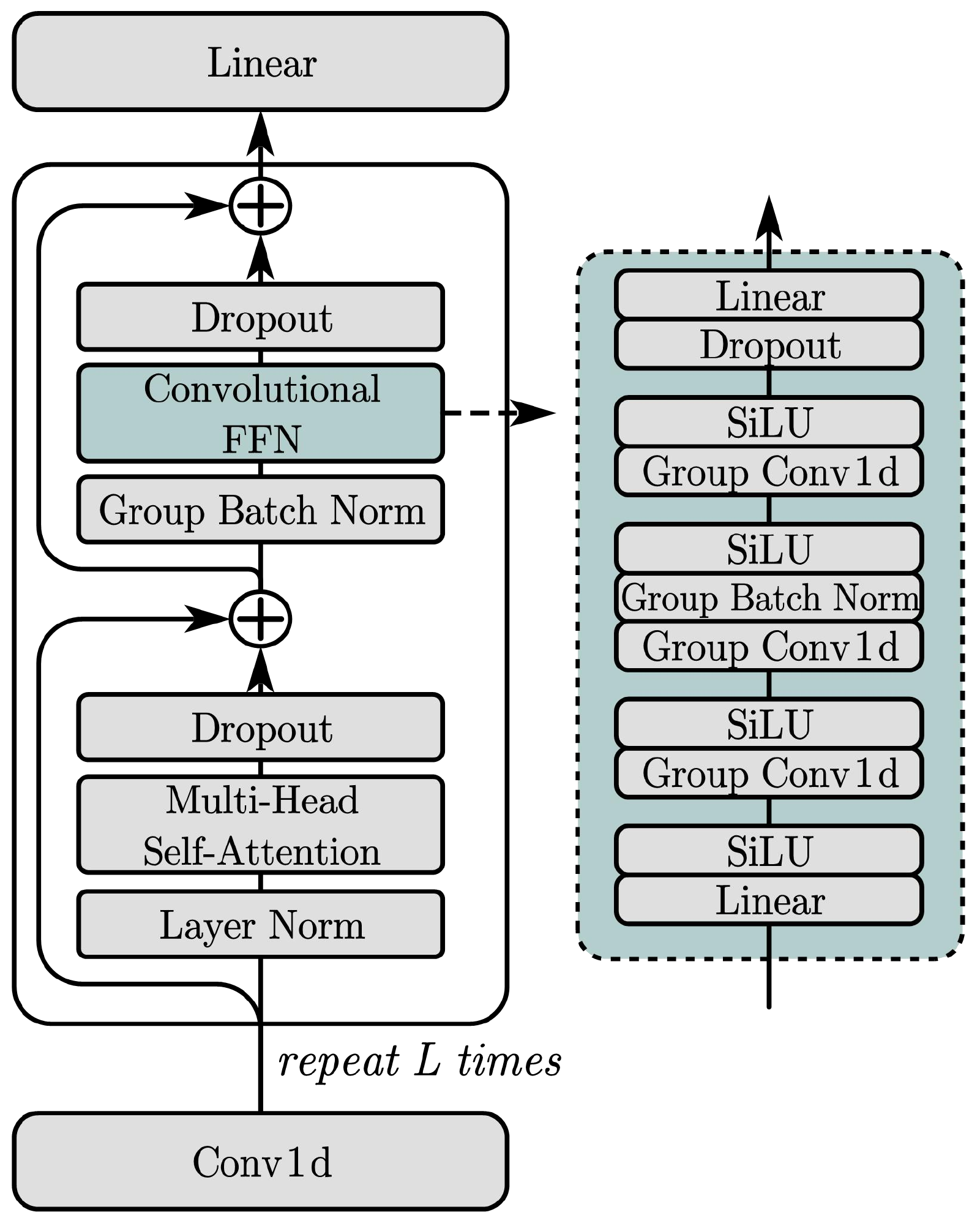}
  \caption{Network architecture of the revised narrow-band Conformer.}
  \label{fig:nbc_v2}
\end{figure}

% 本文提出的网络结构包含三个部分：Conv1d，L层NBC block，以及Linear。
As shown in Fig. \ref{fig:nbc_v2}, the proposed network is composed of one convolutional input layer (Conv1d), $L$ narrow-band conformer (NBC) blocks, and one Linear output layer. The network processes the narrowband signal (as described in Section~\ref{sec:nbss}) layer by layer, and the sequence length for each layer is always $T$. 
As the frequencies are processed independently, we omit the frequency index hereafter.
% Conv1d接受${\rm X}_f$作为输入，输出$x_0[f,:,:] \in R^{T,H_1}$。

Conv1d performs 1-D convolution along the time dimension. It takes ${\rm \textbf{X}}$ as its input sequence, and outputs the sequence of input embedding. The input embedding for one time step is denoted as $\textbf{x}_{0} \in R^{H_1 \times 1}$, where $H_1$ is the number of hidden units.
% 第i个NBC block接收$x_{f,i-1} \in R^{T,H_1}$作为输入，输出$x_{f,i}\in R^{T,H_1}$.
Each NBC block is composed of two parts: the multi-head self-attention module and the convolutional feed forward network (ConvFFN) module, which will be introduced in subsection \ref{sec:MHSA} and \ref{sec:ConvFFN}, respectively.
The output hidden vector of the $l$-th NBC block is denoted as $\textbf{x}_{l}\in R^{H_1 \times 1}$.
% 最后的Linear会接收第L个NBC block的输出然后将其降维得到最后的结果$\widehat{\rm Y}_f$.
The Linear output layer maps the output of the last NBC block to the separated STFT coefficients of different speakers, i.e. $\boldsymbol{\widehat{\rm Y}}$. 
% 因为网络是共享的，因此我们可以认为f属于batch dimension，故而可以使用不含f的表示，即$x_0[:,:]$

\subsection{Multi-head Self-attention Module}% with Sigmoid Weights}
\label{sec:MHSA}

This module consists of a Layer Norm (LN) \cite{ba_LayerNormalization_2016}, a standard Multi-Head Self-Attention (MHSA) \cite{vaswani_attention_2017}, a dropout, and a residual connection from the module input to the module output. This module is formulated as:
\begin{equation}
    \tilde{\textbf{x}}_{l}=\textbf{x}_{l-1}+\text{Dropout}(\text{MHSA}(\text{LN}(\textbf{x}_{l-1}))).
\end{equation}

\subsection{Convolutional Feed Forward Network}
\label{sec:ConvFFN}

This module is composed of one group batch normalization (GBN), one convolutional feed-forward network (ConvFFN), one dropout, and a residual connection. GBN is a new normalization method designed especially for the proposed narrow-band network, and will be introduced later.
The whole module can be formulated as:
\begin{equation}
    \textbf{x}_{l}=\tilde{\textbf{x}}_{l}+\text{Dropout}(\text{ConvFFN}(\text{GBN}(\tilde{\textbf{x}}_{l}))).
\end{equation}

In ConvFFN, a linear layer first transforms the hidden vector from $H_1$-dim to a higher dimension, say $H_2$, then three group convolutional layers are applied, and finally a linear layer transforms the hidden vector from $H_2$-dim back to $H_1$-dim. The SiLU (Sigmoid Linear Unit) \cite{hendrycks_gaussian_2020, ramachandran_searching_2017} activation is applied after the first linear layer and the three convolutional layers. GBN is applied after the second convolutional layer. The group convolutional layers perform 1-D convolution along the time dimension. The number of channels for the convolutional layers is $H_2$, and the channels are split into $G$ groups. 

Compared to the feed forward network used in Transformer \cite{vaswani_attention_2017} and the Conformer network proposed in [15], the major revision of the the proposed ConvFFN is that three convolutional layers are put in between the two linear layers. Using multiple layers of convolutions with a larger number of channels, i.e. $H_2$, accounts for the high requirement of local smoothing/averaging for narrowband speech separation. 

\subsection{Group Batch Normalization}

Group batch normalization (GBN) is especially designed for the proposed narrowband network, and it brings a large performance improvement. As mentioned before, the frequencies of one utterance are processed by the narrowband network separately. 
These separate frequencies are highly correlated according to the common amplitude modulation property \cite{ito_PermutationfreeClusteringRelative_2015}, i.e. the frequency components of one utterance tend to be activated synchronously.
Thus, the frequencies of one utterance can be regarded as a closely correlated group.
Normalizing (the hidden units of) the group members together can better maintain their intrinsic correlation, which can somehow promote the representation capacity of hidden units for different groups.

%Moreover, a complete prediction of the narrow-band network needs the results of all frequencies, thus the frequencies of one utterance are also computationally dependent.
%The strong correlation and computational dependence between training samples makes us to consider to normalize all frequencies of one utterance together, which is the motivation of the proposed entire normalization.

%At training, assume we have $U$ utterances in a mini-batch. 
The hidden units of one network layer is denoted as $h_{u,f,t,i}$, where $u\in\{1,\dots,U\}$, $f\in\{0,\dots,F-1\}$, $t\in\{1,\dots,T\}$ and $i\in\{1,\dots,H\}$ denote the indices of utterance (frequency group) in one mini-batch, frequency (group member), time frame and hidden unit, respectively. Note that $U=1$ at inference.
%As the proposed network processes each frequency separately, the shape of the intermediate hidden units $e$ of one mini-batch containing $T$ frames can be $\mathbb{R}^{UF\times T\times H}$, where $H$ is hidden dimension at one layer, and $UF$ is total number of training samples for the proposed network for this mini-batch ($U$ utterances and $F$ frequencies per utterance).
%For the convenience of description, let's reshape $e$ to $\mathbb{R}^{U\times F\times T\times H}$.
% 对其做Frequency normalization就可以表示为：
GBN can be formulated as:
\begin{align}
    \text{GBN}(h)_{u,f,t,i} = & \frac{h_{u,f,t,i}-\mu_{u,t}}{\sqrt{\sigma_{u,t}^2+\epsilon}}\gamma_i+\beta_i,
\end{align}
where
\begin{align}
    \mu_{u,t} = & \frac{1}{FH} \sum_{f,i} h_{u,f,t,i}, \label{eq_mu}\\
    \sigma_{u,t}^2 = & \frac{1}{FH} \sum_{f,i} (h_{u,f,t,i}-\mu_{u,t})^2, \label{eq_delta}
\end{align}
and $\epsilon$ is a small value for computational stability, $\gamma_i, \ \beta_i, \ i\in\{1,\dots,H\}$ are learnable parameters. 
Here the mean and variance are not calculated over all time steps, as our preliminary experiments show that normalization over time steps will degrade the performance.

% 从mu_u_t的定义中可以看出，同一句子不同频带的hidden units会共享同一个均值和标准差。
We can see that, besides the hidden units of one layer (as is done in LN), all frequencies of one utterance also share the same mean and variance.
% 这种共享会使得不同频带的分布差异得以保留，我们认为这就是EN取得巨大收益的原因。
This keeps the distribution of hidden units over all frequencies unchanged after the normalization, which we think is the major advantage of the proposed GBN. GBN is similar with BN, as different frequencies are processed independently and considered as independent samples during training. However, different from BN, the proposed network can apply the same GBN for both training and inference, as all frequencies of one utterance also present at inference. 
By contrast, BN uses the statistics of one mini-batch during training, and the moving average of the statistics calculated at training for inference, which may harm the performance.

\section{Experimental Setup}
\subsection{Datasets}
The proposed method is evaluated on two spatialized versions of the WSJ0-2mix dataset \cite{hershey_DeepClusteringDiscriminative_2016}, and both for the two-speaker separation task. WSJ0-2mix contains 20,000, 5,000 and 3,000 speech pairs for training, validation and test, respectively.
One dataset is developed in this paper using simulated room impulse responses with an 8-channel circular microphone array.
The other was proposed in \cite{wang_multi-channel_2018} with simulated and randomly sampled microphone geometries, and is used in some recent works \cite{ochiai_beam-tasnet_2020, chen_BeamGuidedTasNetIterative_2022}.
For simplicity, we call the former as circular array dataset, and the later random array dataset.

\subsubsection{Circular array dataset}
\begin{figure}[t]
  \centering
  \includegraphics[width=0.8\linewidth]{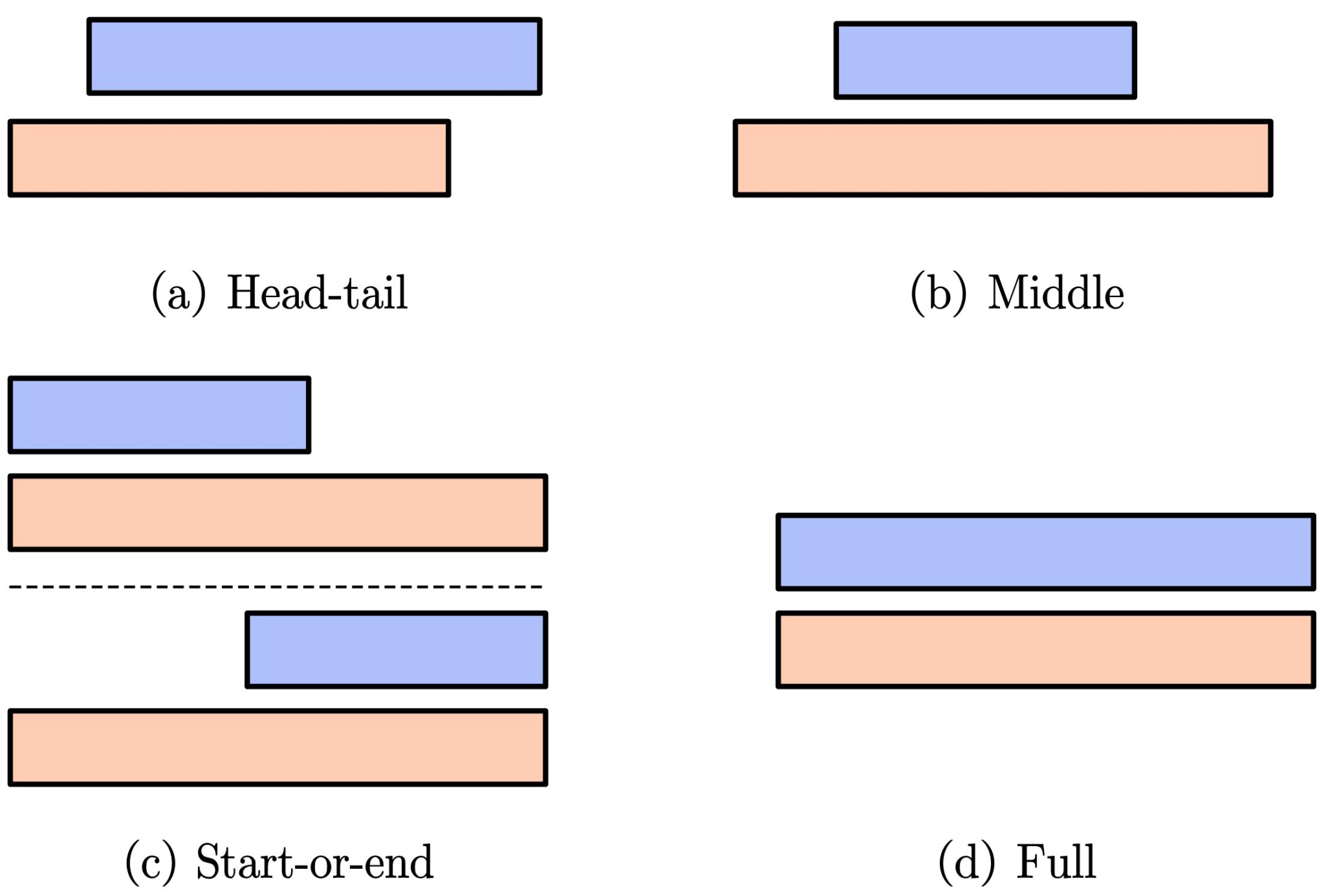}
  \caption{Four overlap ways. Different colors represent different speakers.}
  \label{fig:ovlps}
\end{figure}

Normally, one speech pair is not fully overlap in practice. To account for this, each speech pair is mixed with an overlap ratio being uniformly sampled from the range of [10\%, 100\%]. For speech separation, the non-overlapped segments may also be important, from which some separation features of individual speakers can be extracted. Thence, the overlap way that determines the position of non-overlapped portions may affect the performance of speech separation. Fig. \ref{fig:ovlps} shows four overlap ways that may happen in practice and will be considered in this work.  
Each speech pair is overlapped with all the four ways. 

The sample rate is 16 kHz. The mixed utterances are all set to four-second long.
For the utterances shorter or longer than 4 seconds, they will be concatenated with other utterances of the same speaker, or randomly clipped to four seconds.
Room impulse responses are simulated using a GPU based implementation of the image method \cite{allen_image_1979}, called gpuRIR \cite{diaz-guerra_gpurir_2021}.
The length, width and height of the simulated rooms are uniformly sampled in the range of [3 m, 8 m], [3 m, 8 m] and [3 m, 4 m], respectively.
The reverberation time (RT60) of each room is uniformly sampled in the range of [0.1 s, 1.0 s].
A horizontal 8-channel circular microphone array with a radius of 5 cm is used.
The center of microphone array is randomly put in a square area (diameter is 1 m) at the room center with a height of 1.5 m. 
Speaker locations are randomly sampled in the room with a height of 1.5 m and with the direction difference between two speakers randomly sampled from 0$^\circ$ to 180$^\circ$.
Each speaker is located at least 0.5 m away from the walls.
The signal of each speaker is re-scaled such that the signal-to-noise ratio (SNR) for each speaker with respect to the other speaker is uniformly distributed in $[-5,5]$ dB. 

\subsubsection{Random array dataset}
This dataset was proposed in \cite{wang_multi-channel_2018}.
It uses 8-channel microphone arrays with aperture sizes sampled randomly from 15 cm to 25 cm.
% For fair comparison with \cite{ochiai_beam-tasnet_2020, chen_BeamGuidedTasNetIterative_2022}, we use the first four channels.
The rooms are randomly generated, and the corresponding room impulse responses are simulated with RT60 randomly drawn from 0.2 s to 0.6 s.
The speech pairs are overlapped in the \emph{full} manner or in the \emph{start} manner shown in Fig. \ref{fig:ovlps}, referred to respectively as ``min"-type and ``max"-type in \cite{hershey_DeepClusteringDiscriminative_2016, wang_multi-channel_2018}.
SNR also uniformly distributes in $[-5,5]$ dB. The sample rate could be either 8 kHz or 16 kHz.

\subsection{Training Configurations}

For the proposed network, we set the kernel size of the convolutional input layer to 5. The kernel size and the number of groups of group convolutions are set to 3 and 8, respectively.
A small version and a large version of the proposed network are proposed/suggested. 
The small network, referred to as NBC2-small, uses 8 NBC blocks and 2 attention heads; and the numbers of hidden units are set to $H_1=96$ and $H_2=192$.
The large network, referred to as NBC2-large, uses 12 NBC blocks and 2 attention heads; and the numbers of hidden units are set to $H_1=192$ and $H_2=384$. Their model sizes are 0.9 M and 5.6 M, respectively. 

STFT is applied using a Hanning window with a length of 512/256 samples (32ms) and a hop size of 256/128 samples for the 16/8 kHz data. 
The network is trained with 2 utterances per mini-batch, thus the batch size of narrow-band training samples is 514 ($2 \times 257$) frequencies.
The Adam \cite{kingma2015adam} optimizer is used with a learning rate initialized to 0.001 and  exponentially decayed as $lr \xleftarrow{} 0.001 * 0.99^{epoch}$.
Gradient clipping is applied with a gradient norm threshold of 5.
For all experiments, the proposed network is trained for 100 epochs.

\subsection{Performance Metrics}
We evaluate the proposed network using the metrics of 1) the number of parameters, 2) the real time factor (RTF) tested on a personal computer equipped with Intel(R) i7-9700 CPU (3.0 GHz), and 3) the speech separation performance in terms of perceptual evaluation of speech quality (PESQ) \cite{rix_perceptual_2001} and signal-to-distortion ratio (SDR) \cite{vincent_performance_2006, scheibler_SDRMediumRare_2022}), where narrow-band PESQ and wide-band PESQ are used for the 8 kHz and 16 kHz data, respectively. 

\begin{table*}[htb]
\setlength\tabcolsep{4pt}
\renewcommand{\arraystretch}{1.3}
\caption{Separation performance on circular array dataset. WB-PESQ and SDR (dB) are reported in the format of "WB-PESQ/SDR".}
\label{table_circular_dataset}
\centering
\begin{tabular}{cccc|ccccc}
\hline\hline
\multirow{2}{*}{Method} & \multirow{2}{*}{\#CHN} & \multirow{2}{*}{\#Param} & \multirow{2}{*}{RTF} & \multicolumn{5}{c}{\textbf{Performance} (WB-PESQ / SDR (dB))}\\
 & & & & Head-tail & Middle & Start-or-end & Full & \textbf{Average} \\
\hline
\multirow{1}{*}{unproc.} & 1/2/4/8 & - & - & 1.60 / \ 0.2 & 1.64 / \ 0.5 & 1.68 / \ 0.4 & 1.31 / \ 0.2 & 1.56 / \ 0.3 \\
\hline
\multirow{3}{*}{FaSNet-TAC \cite{luo_end--end_2020}} & 2 & 2.8 M & 0.27 & 2.32 / 11.9 & 2.18 / 10.6 & 2.24 / 11.0 & 1.86 / \ 9.0 & 2.15 / 10.6 \\
 & 4 & 2.8 M & 0.45 & 2.49 / 13.0 & 2.33 / 11.7 & 2.39 / 12.1 & 2.04 / 10.1 & 2.31 / 11.7 \\
 & 8 & 2.8 M & 0.73 & 2.54 / 13.2 & 2.38 / 11.9 & 2.45 / 12.4 & 2.09 / 10.3 & 2.37 / 12.0 \\
\hline
\multirow{2}{*}{Beam-Guided TasNet \cite{chen_BeamGuidedTasNetIterative_2022} (iter=2)} & 2 & 5.3 M & 0.50 & 2.88 / 15.4 & 2.66 / 13.9 & 2.70 / 14.1 & 2.36 / 12.1 & 2.65 / 13.9 \\
 & 4 & 5.7 M & 0.61 & 3.12 / 17.4 & 2.83 / 15.4 & 2.84 / 15.5 & 2.50 / 13.4 & 2.82 / 15.4 \\
\hline
\multirow{4}{*}{SepFormer \cite{subakan_attention_2021}} & 1 & 25.7 M & 1.57 & 2.71 / 13.5 & 2.54 / 12.5 & 2.62 / 12.9 & 2.26 / 10.6 & 2.53 / 12.4 \\
 & 2 & 25.7 M & 1.51 & 3.06 / 15.4 & 2.85 / 14.2 & 2.93 / 14.7 & 2.66 / 12.5 & 2.88 / 14.2 \\
 & 4 & 25.7 M & 1.51 & 3.22 / 16.2 & 3.00 / 15.1 & 3.07 / 15.5 & 2.85 / 13.3 & 3.03 / 15.0 \\
 & 8 & 25.7 M & 1.51 & 3.20 / 16.3 & 2.99 / 15.2 & 3.07 / 15.6 & 2.84 / 13.4 & 3.03 / 15.1 \\
\hline
\multirow{3}{*}{NB-BLSTM \cite{quan_MultiChannelNarrowBandDeep_2022}} & 2 & 1.2 M & 0.39 & 2.43 / 11.2 & 2.16 / \ 9.7 & 2.20 / \ 9.8 & 1.83 / \ 8.1 & 2.16 / \ 9.7 \\
 & 4 & 1.2 M & 0.40 & 2.45 / 11.4 & 2.29 / 10.4 & 2.31 / 10.6 & 2.02 / \ 9.3 & 2.27 / 10.4 \\
 & 8 & 1.2 M & 0.40 & 2.78 / 13.0 & 2.55 / 11.6 & 2.58 / 11.9 & 2.35 / 10.5 & 2.56 / 11.8 \\
\hline
\multirow{3}{*}{NBC2-small (prop.)} & 2 & 0.9 M & 0.96 & 3.71 / 19.1 & 3.43 / 17.4 & 3.48 / 17.8 & 3.35 / 15.9 & 3.49 / 17.5 \\
 & 4 & 0.9 M & 0.97 & 3.89 / 21.2 & 3.62 / 19.2 & 3.66 / 19.6 & 3.54 / 17.7 & 3.68 / 19.4 \\
 & 8 & 0.9 M & 0.97 & 3.87 / 21.0 & 3.59 / 19.0 & 3.64 / 19.4 & 3.57 / 17.7 & 3.67 / 19.3 \\
\hline
\multirow{3}{*}{NBC2-large (prop.)} & 2 & 5.6 M & 2.78 & 4.01 / 21.4 & 3.76 / 19.5 & 3.80 / 19.9 & 3.74 / 18.1 & 3.83 / 19.7 \\
 & 4 & 5.6 M & 2.77 & 4.18 / 24.3 & 3.95 / 21.9 & 3.97 / 22.4 & 3.97 / 20.8 & 4.02 / 22.3 \\
 & 8 & 5.6 M & 2.77 & 4.22 / 25.2 & 3.99 / 22.6 & 4.01 / 23.1 & 4.01 / 21.4 & 4.06 / 23.1 \\
\hline\hline
\end{tabular}
\end{table*}

\section{Experiments and Discussions}

\subsection{Results on Circular Array Dataset}
On the circular array dataset, we compare the proposed method with the following methods, for which publicly released code is available and the network is trained from scratch using the circular array dataset.

\begin{itemize}[leftmargin=*]
    \item FaSNet-TAC \cite{luo_end--end_2020}: A filter-and-sum network with transform-and-concatenate mechanism. % 说下训练的策略？
    \item SepFormer \cite{subakan_attention_2021}: A transformer-based single channel speech separation model. SepFormer is modified in our experiments to account for the multichannel case, by simply changing the input channel of its first convolution layer from 1 to the number of microphone channels. Note that this may not be the optimal way to extend SepFormer to the multichannel case, but it still improves the performance when increasing the number of channels. 
    \item Beam-Guided TasNet \cite{chen_BeamGuidedTasNetIterative_2022}: A two-stage neural beamformer. In the first stage, a neural network is used to estimate the multichannel speech signal of each speaker; while in the second stage, another neural network is applied to iteratively fine-grain the separated multichannel speech signals and apply MVDR beamforming. 
    Beam-Guided TasNet was trained with 8 kHz signals in its original paper, we doubled the kernel size and stride of its filterbank layer for this 16 kHz dataset, as advised by its own authors.
    % \item Oracle MVDR\footnote{https://github.com/Enny1991/beamformers}: Estimates the steering vector of target speaker by using the covariance matrices of ground truth speech signals.
    \item NB-BLSTM \cite{quan_MultiChannelNarrowBandDeep_2022}: Our previously proposed narrow-band speech separation method. Instead of using the Conformer network proposed in this work, it uses two layers of bidirectional LSTM network.
\end{itemize}

In Table \ref{table_circular_dataset}, the results are reported for the four overlap ways. The results of using a sub-array with two or four microphones are also reported, where the sub-arrays are uniformly selected from the 8-channel array. The 2-, 4- and 8-channel arrays take the same  reference channel. 
%The 2-channel (0-th and 4-th channels), 4-channel (0-th, 2-th, 4-th and 6-th channels), and 8-channel results of each algorithm are reported, except Beam Guided TasNet.
For Beam-Guided TasNet, the 8-channel results are not reported, as it couldn't obtain reasonable performance in our experiments.
The single-channel performance of SepFormer is also reported, as it's originally proposed for single-channel speech separation.
%For the proposed method, two networks with different configurations are reported, i.e. NBC2-small and NBC2-large. The configurations of the two networks can be found in Table \ref{table_parameters}.

From Table \ref{table_circular_dataset}, we can see that the performance of FasNet-TAC and our previously proposed NB-BLSTM are comparable, which indicates the effectiveness of narrow-band speech separation even with a simple BLSTM network. Beam-Guided TasNet and SepFormer notably outperform FasNet-TAC and NB-BLSTM. The success of SepFormer verifies that the self-attention mechanism is especially fit for speech separation, as clustering the frames of different speakers is one foundation for separating speakers. For speaker clustering, single-channel SepFormer may rely on some speaker features, while multichannel SepFormer may also rely on the spatial features as employed by the proposed narrow-band network. The proposed NBC2-small outperforms comparison methods by a large margin, by using only 0.9 M parameters. NBC2-large further largely improves the performance. This demonstrates that the narrow-band spatial information is highly discriminative for speech separation. The proposed method is effective to fully leverage this information by 1) setting a dedicated narrow-band network to focus on this information, and 2) leveraging a powerful improved conformer network. The computational complexity and thus RTF of the proposed network are actually relative high, since each frequency needs to run the network one time. NBC2-small basically reaches the real time requirement. 

Among the four overlap ways, the \emph{full} overlap case achieves the worst performance, as it is the most difficult case due to the 100\% overlap ratio. The performance of \emph{head-tail} is the best. The performance of \emph{middle} and \emph{start-or-end} are comparable and worse than the one of \emph{head-tail}. The reason is: non-overlapped segments are important for extracting  separation features of individual speakers. \emph{head-tail} has non-overlapped segments for both the two speakers, while \emph{middle} and \emph{start-or-end} only for one of the two speakers. 

For the two-speaker separation problem, using more microphones can generally promote the separation performance.
As shown in Table \ref{table_circular_dataset}, the performance promotion is especially obvious for all the methods when increasing the number of microphones from two to four. 
However, it becomes less significant when increasing the number of microphones further to eight, which indicates that the spatial information provided by four microphones is sufficiently discriminative for separating two speakers.

\subsection{Results on Random Array Dataset}
For the random array dataset \cite{wang_multi-channel_2018}, we compare the proposed method with MC (multi-channel) Deep Clustering \cite{wang_multi-channel_2018}, the \emph{Parallel Encoder} in \cite{gu_EndtoEndMultiChannelSpeech_2019} (refered to as MC-TasNet following \cite{ochiai_beam-tasnet_2020}), Beam-TasNet \cite{ochiai_beam-tasnet_2020}, Beam-Guided TasNet  \cite{chen_BeamGuidedTasNetIterative_2022} and oracle MVDR. The results are directly quoted from the related papers, as explained in Table \ref{table_random_array}.
Except that, for full comparison, we retrained Beam-Guided TasNet (iter=2) \cite{chen_BeamGuidedTasNetIterative_2022} for both the 8 kHz and 16 kHz cases. Note that, the SDR  score of our retrained model for the 8 kHz case is slightly better than the one reported in its original paper.
Both Beam-Guided TasNet and our proposed networks are trained with the ``min"-type data (the \emph{full} overlap shown in Fig. \ref{fig:ovlps}). Experiments are conducted using the first four channels.

\begin{table}[tpb]
\setlength\tabcolsep{2pt}
\renewcommand{\arraystretch}{1.2}
\caption{Four-channel speech separation results on the random array dataset. $^{\star}$, $^{\ast}$ and $^{\dagger}$ denote that the scores are quoted from \cite{wang_multi-channel_2018}, \cite{ochiai_beam-tasnet_2020} and \cite{chen_BeamGuidedTasNetIterative_2022}, respectively. 
}
\label{table_random_array}
\centering
\resizebox{\linewidth}{!}{
\begin{tabular}{ccccc}
\toprule
\multirow{2}{*}{\textbf{Method}} & \multicolumn{2}{c}{8k} & \multicolumn{2}{c}{16k}\\
~ & NB-PESQ & SDR (dB) & WB-PESQ & SDR (dB) \\
\midrule
unproc. & 2.00 & 0.1 & 1.45 & 0.1 \\
\hline
% MC Deep Clustering & 2 & 8.9 & - \\
%  & 4 & 9.4 & - \\
MC Deep Clustering \cite{wang_multi-channel_2018} & - & 9.4$^{\star}$ & - & - \\
% \hline
% TasNet & 1 & - & - & - & 11.3 \\
%\hline
% MC-TasNet & 2 & - & 12.7 \\
%  & 4 & - & 12.1 \\
MC-TasNet \cite{gu_EndtoEndMultiChannelSpeech_2019}  & - & - & - & 12.1$^{\ast}$ \\
%\hline
% BeamTasNet & 2 & - & 13.8 \\
%  & 4 & 17.4 & 16.8 \\
Beam-TasNet \cite{ochiai_beam-tasnet_2020} & - & 17.4$^{\dagger}$ & - & 16.8$^{\ast}$ \\
%Oracle signal MVDR & - & - & - & 21.7$^{\ast}$ \\
%\hline
Beam-Guided TasNet \cite{chen_BeamGuidedTasNetIterative_2022} (iter=2) & 3.90 & 20.8 & 3.48 & 19.0 \\
Beam-Guided TasNet \cite{chen_BeamGuidedTasNetIterative_2022} (iter=4) & - & 21.5$^{\dagger}$ & - & - \\
Oracle signal MVDR & - & 23.5$^{\dagger}$ & - & 21.7$^{\ast}$ \\
\hline
% SepFormer & 4 & 3.6 / 15.6 & - \\
% \hline
NBC2-small (prop.) & 4.14 & 22.3 & 4.03 & 22.2\\
% NBC2-base & & & \\
NBC2-large (prop.) & 4.31 & 26.1 & 4.28 & 26.2 \\
\bottomrule
\end{tabular}
}
\end{table}

Table \ref{table_random_array} shows the speech separation results of the ``max"-type data (the \emph{start} overlap shown in Fig. \ref{fig:ovlps}).
We can see that Beam-TasNet shows the superiority of beamforming over the binary-mask-based method (i.e. MC Deep Clustering), and the end-to-end time-domain method (i.e. MC-TasNet).
Beam-Guided TasNet further improves the performance of Beam-TasNet by iteratively refining the  separated results, at the cost of larger training and inference time.
The proposed networks outperform these comparison methods. Especially, NBC2-large even notably outperforms the oracle MVDR, while the latter can be regarded as the upper bound of beamforming-based methods.
%The results of NBC2-small are better than other algorithms.
%NBC2-large further improves the separation performance even better the oracle MVDR results, i.e. 25 dB vs. 23.5 dB for 8 kHz sample rate reported in Beam Guided TasNet \cite{chen_BeamGuidedTasNetIterative_2022} and 24.7 dB vs. 21.7 dB for 16 kHz sample rate reported in Beam TasNet \cite{ochiai_beam-tasnet_2020}, which can be regarded as the upper bound of Beam Guided TasNet and Beam TasNet.

Compared with the circular array dataset, this random array dataset is simpler to process, as both Beam-Guided TasNet and the proposed networks achieve better performance measures on this dataset.
This is reasonable as the circular array dataset has larger reverberation than this dataset, with the RT60 of [0.1, 1.0] s versus [0.2, 0.6] s. Although we don't perform dereverberation, it is still important to properly model the reverberation effect for reverberant speech separation.  
Beam-Guided TasNet and the proposed method perform well on this random array dataset, which  demonstrates the generalization ability of the two methods in terms of array variation.

\subsection{Ablation Studies}
On the 8-channel circular array dataset, we conduct ablation experiments of the proposed NBC2 network. 
Table \ref{table_parameters} shows the results.
In this table, we first report the performance of a \emph{base} network.
Then ablation experiments are conducted by changing some hyperparameters of the \emph{base} network.
% For simplicity, we omit the configurations in other lines same with the \textit{base} network.

\begin{table*}[htpb]
\setlength\tabcolsep{4pt}
\renewcommand{\arraystretch}{1.3}
\caption{Ablation experiments on the 8-channel circular array dataset. BN$^\dag$ is trained with a mini-batch size of 8 utterances.}
\label{table_parameters}
\centering
\begin{tabular}{c|ccccccc|cccccc}
\hline
 & \multicolumn{7}{c|}{\textbf{Network Parameters}} & \multicolumn{5}{c}{\textbf{Performance} (WB-PESQ / SDR (dB))}\\
 & \#Block & \#Head & $H_1$ & $H_2$ & Normalization & \#Param & RTF & Head-tail & Middle & Start-or-end & Full & \textbf{Average} \\
\hline
base & 8 & 2 & 128 & 256 & GBN & 1.7 M & 1.20 & 4.04 / 22.8 & 3.77 / 20.5 & 3.81 / 20.9 & 3.78 / 19.4 & 3.85 / 20.9 \\
\hline
\multirow{4}{*}{(\textit{A})} & 8 & 2 & 128 & 256 & BN & 1.7 M & 0.91 & 3.16 / 16.2 & 2.88 / 14.0 & 2.90 / 14.3 & 2.81 / 14.0 & 2.94 / 14.6 \\
 & 8 & 2 & 128 & 256 & BN$^\dag$ & 1.7 M & 0.91 & 3.30 / 17.0 & 3.03 / 15.1 & 3.05 / 15.3 & 2.93 / 14.4 & 3.08 / 15.5 \\
 & 8 & 2 & 128 & 256 & LN & 1.7 M & 0.92 & 3.49 / 18.8 & 3.19 / 16.7 & 3.21 / 16.9 & 3.14 / 16.0 & 3.26 / 17.1 \\
 & 8 & 2 & 128 & 256 & GN & 1.7 M & 0.93 & 3.54 / 19.4 & 3.27 / 17.3 & 3.28 / 17.6 & 3.19 / 16.4 & 3.32 / 17.7 \\
\hline
\multirow{2}{*}{(\textit{B})} & 8 & 4 & 128 & 256 & GBN & 1.7 M & 1.28 & 4.04 / 23.1 & 3.77 / 20.8 & 3.80 / 21.2 & 3.78 / 19.6 & 3.85 / 21.2 \\
 & 8 & 8 & 128 & 256 & GBN & 1.7 M & 1.52 & 4.06 / 23.0 & 3.80 / 20.8 & 3.83 / 21.2 & 3.82 / 19.6 & 3.88 / 21.2 \\
\hline
\multirow{6}{*}{(\textit{C})} & 8 & 2 & 32 & 64 & GBN & 0.1 M & 0.35 & 3.30 / 16.7 & 3.03 / 15.2 & 3.09 / 15.6 & 2.89 / 13.7 & 3.08 / 15.3 \\
 & 8 & 2 & 64 & 128 & GBN & 0.4 M & 0.62 & 3.73 / 19.9 & 3.47 / 18.2 & 3.51 / 18.6 & 3.31 / 16.5 & 3.51 / 18.3 \\
\hspace{1.2cm} \textbf{small} & \textbf{8} & \textbf{2} & \textbf{96} & \textbf{192} & \textbf{GBN} & \textbf{0.9 M} & \textbf{0.97} & \textbf{3.87} / \textbf{21.0} & \textbf{3.59} / \textbf{19.0} & \textbf{3.64} / \textbf{19.4} & \textbf{3.57} / \textbf{17.7} & \textbf{3.67} / \textbf{19.3} \\
 & 8 & 2 & 192 & 384 & GBN & 3.7 M & 1.73 & 4.12 / 23.9 & 3.87 / 21.4 & 3.90 / 21.9 & 3.87 / 20.2 & 3.94 / 21.9 \\
 & 8 & 2 & 64 & 256 & GBN & 1.0 M & 1.00 & 3.89 / 21.2 & 3.61 / 19.1 & 3.65 / 19.5 & 3.57 / 17.8 & 3.68 / 19.4 \\
 & 8 & 2 & 192 & 768 & GBN & 8.9 M & 2.94 & 4.18 / 24.6 & 3.93 / 22.1 & 3.96 / 22.5 & 3.96 / 21.0 & 4.01 / 22.6 \\
\hline
\multirow{2}{*}{(\textit{D})} & 12 & 2 & 128 & 256 & GBN & 2.5 M & 1.72 & 4.16 / 24.1 & 3.91 / 21.7 & 3.94 / 22.1 & 3.91 / 20.4 & 3.98 / 22.1 \\
\hspace{1.2cm} \textbf{large} & \textbf{12} & \textbf{2} & \textbf{192} & \textbf{384} & \textbf{GBN} & \textbf{5.6 M} & \textbf{2.77} & \textbf{4.22} / \textbf{25.2} & \textbf{3.99} / \textbf{22.6} & \textbf{4.01} / \textbf{23.1} & \textbf{4.01} / \textbf{21.4} & \textbf{4.06} / \textbf{23.1} \\
\hline
\end{tabular}
\vspace{-0.2cm}
\end{table*}

\subsubsection{Group Batch Normalization}
In the proposed NBC2 network, there are two GBN layers in each NBC block.
We conducted ablation experiments in group (\textit{A}) to replace the GBN layer before ConvFFN with LN following the configuration of Transformer \cite{vaswani_attention_2017} and Conformer \cite{gulati2020conformer}, and replace the GBN layer after the second convolutional layer with either BN, LN, or GN.
For BN, we also trained one network, denoted as BN$^\dag$, with a mini-batch size of 8 utterances. 
The number of groups of GN is set to 8, according to the configuration of group convolutional layers.
BN is widely used for convolutional layers in the literature, including the convolutional layers of Conformer \cite{gulati2020conformer}, which however performs badly in this experiment, due to the small mini-batch size, i.e. two utterances per mini-batch. Although the frequencies of one utterance are considered as independent training samples, they don't effectively enlarge the mini-batch size as they are highly correlated. By improving the mini-batch size to 8 utterances, BN$^\dag$ improves the performance, at the cost of a very large memory consumption for $8\times257$ training samples. For this small mini-batch case, LN and GN notably outperforms BN. GN is more suitable than LN for the group convolutional layers used in the proposed network. 
%As we can see, increasing the number of training samples of BN to $8\times 257$ can improve the performance of BN, but BN still largely underperforms LN.
%GN performs better LN, while the proposed GBN performs much better than GN.
The superior performance of GBN shows its effectiveness for the proposed narrow-band network.

\subsubsection{The number of attention heads}
The results in (\textit{B}) show that using 4 or 8 attention heads provides little performance gain compared with using 2 heads, but requires a larger inference time, i.e. RTF. 
Moreover, more attention heads require more memory and training time. 
Hence, we use 2 attention heads for the proposed network.

\subsubsection{Hidden Dimensions}
% + v216 v217 得到hidden扩到四倍的结果
The results in (\textit{C}) show the performance of the proposed network using different number of hidden units, i.e. $H_1$ and $H_2$. We test two configurations, i.e. $H_2=2\times H_1$ (the first four rows) and $H_2=4\times H_1$ (the last two rows). 
%When $H_2=2\times H_1$, i.e. \textit{base} and the first four rows in \textit{(C)},
It can be seen that increasing the dimension of hidden units constantly increase the performance of the proposed network. We find that setting $H_2=2\times H_1$ is better than setting $H_2=4\times H_1$, as the former has a higher parameter efficiency. For example, compared with the fifth row, the third row achieves similar performance measures with a smaller model. 
%We have also tested the cases where $H_2=4\times H_1$, i.e. 5th and 6th rows in \textit{(B)}.
%Comparing the experiments of the two settings, we can know 1) increasing $H_2$ can indeed promote the performance and 2) however if we compare the two settings under similar model size, i.e. the 3rd row versus 5th row and 8th row versus 6th row in \textit{(B)}, the setting $H_2=4\times H_1$ doesn't give too much benefit over $H_2=2\times H_1$.
Therefore, we use the configuration of $H_2=2\times H_1$ for the proposed network.

\subsubsection{Number of blocks}
In (\textit{D}), the number of NBC blocks is increased to 12, which improves the performance measures accordingly. Relative to the the \emph{base} network, when comparing the fourth row of (\textit{C}) and the first row of (\textit{D}), we can see that increasing the number of layers seems more parameter efficient than increasing the dimension of hidden units.

\subsubsection{The Proposed/suggested Small and Large Networks}
According to the speech separation performance, model size, and training/inference speed, we propose to use the \emph{small} and \emph{large} networks as shown (in bold) in Table \ref{table_parameters}.

\subsection{Spectrum-Agnostic Experiment}
\label{sec_Spectra_Agnostic}

The proposed method processes frequencies independently, and it does not learn any knowledge about spectral patterns. Thence it should be agnostic to the full-band speech spectra. To verify this, we retrained the proposed network and the comparison networks using the data of only four speakers, to see whether the networks can generalize well to unseen speakers with a limited number of training speakers. This experiment is conducted on the 4-channel circular array dataset.
The four speakers, including two males and two females, are randomly selected from the original training set. The 120 longest clean utterances of each speaker are used, and the total time duration of selected utterances is about one hour. The utterances for different speakers are randomly mixed, and a total of 86,400 speech pairs are generated.  The test set is kept unchanged. 
%Compared with the original 4-channel training set, all settings are the same except for the number of speakers and the number of speech pairs (86400 speech pairs in this training set, while 20000 speech pairs in the original training set).

Table \ref{table_4chn_4spk} reports the results, where ``full training" means the networks are trained with the original training set.    
%performance of each algorithm on the 4-channel circular array test set trained with the 4-speaker train set (the ``4 speakers" column) and trained with the original train set (the ``original" column).
From this table, we can see that with limited training speakers and utterances, FaSNet-TAC, Beam-Guided TasNet and SepFormer all have the spectral generalization problem, as the performance measures of them all degrade from ``full training" to ``4-speaker training". 
Our narrow-band methods, i.e. NB-BLSTM and NBC2-small, do not suffer from the spectral generalization problem as we expected.
The performance of NB-BLSTM and NBC2-small with ``4-speaker training" is even better than the ones with ``full training", which is possibly due to the increased number of training speech pairs (86,400 versus 20,000 pairs).
A good spectral generalization capability is important for real applications, as the proposed network can be easily trained with a very limited amount of training data, e.g. one hour of clean utterances in this experiment. 

\begin{table}[tbp]
\setlength\tabcolsep{2pt}
\renewcommand{\arraystretch}{1.3}
\caption{Spectral generalization experiment.}
\label{table_4chn_4spk}
\centering
\begin{tabular}{c|cc|cc}
\hline
\multirow{2}{*}{Method} & \multicolumn{2}{c|}{full training} & \multicolumn{2}{c}{4-speaker training} \\
 & WB-PESQ & SDR  & WB-PESQ & SDR \\
 & & (dB) & & (dB) \\
% Method & \multicolumn{2}{c}{\textbf{Performance}}\\
\hline
FaSNet-TAC \cite{luo_end--end_2020} & 2.31 & 11.7 & 2.19 & 10.4 \\
\hline
Beam-Guided TasNet \cite{chen_BeamGuidedTasNetIterative_2022} (iter=2) & 2.82 & 15.4 & 2.56 & 13.2 \\
\hline
SepFormer \cite{subakan_attention_2021} & 3.03 & 15.0 & 2.75 & 13.8 \\
\hline
NB-BLSTM \cite{quan_MultiChannelNarrowBandDeep_2022} & 2.27 & 10.4 & 2.34 & 10.8 \\
\hline
NBC2-small (prop.) & {3.68} & {19.4} & {3.72} & {19.9} \\
\hline
\end{tabular}
\end{table}

\subsection{Attention Analysis}

\begin{figure*}[htbp]
  \centering
  \includegraphics[width=1\linewidth]{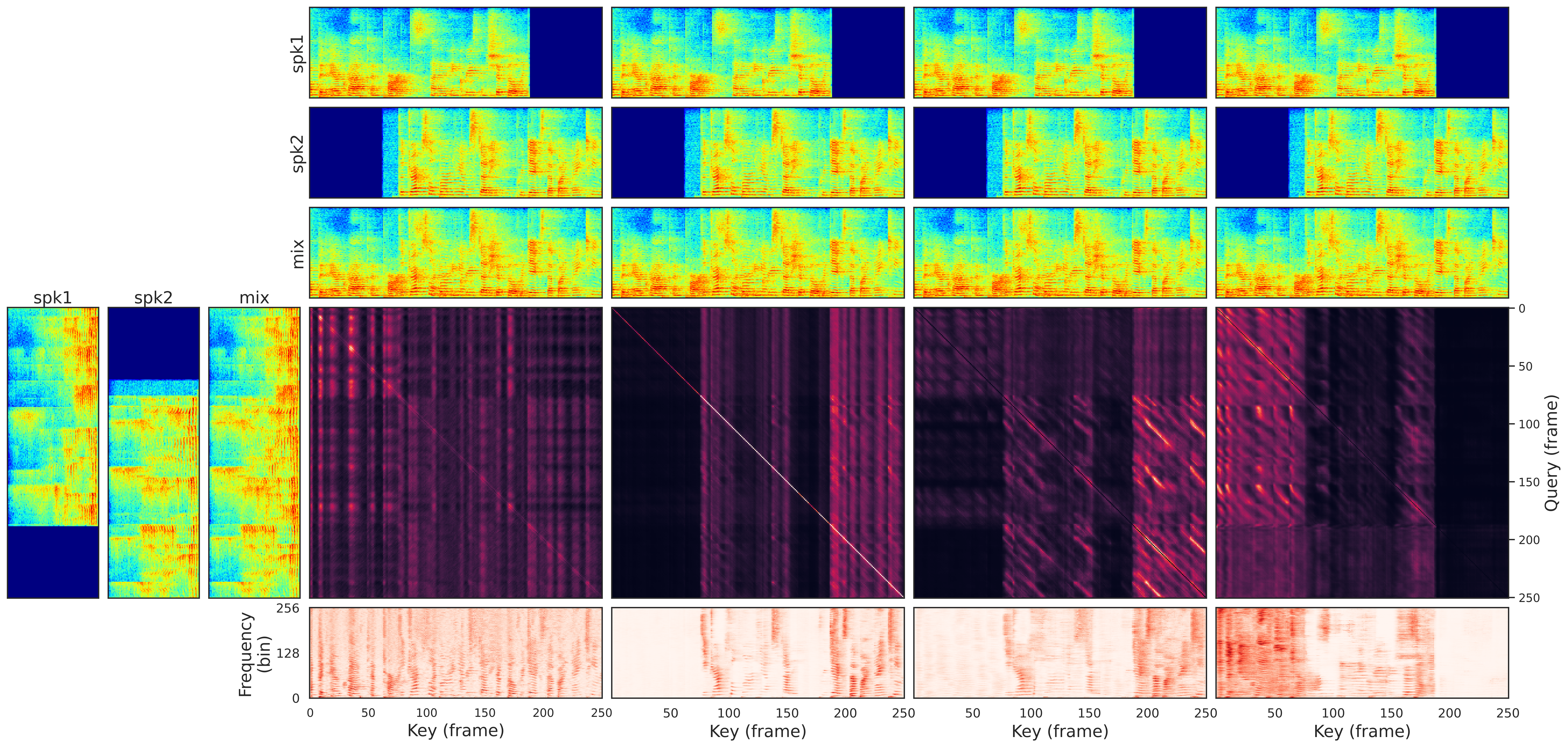}
  \caption{Attention maps of one utterance. RT60 is 0.8 s. `spk1' and `spk2' stand for the speech signal of the first and second speakers, respectively, while `mix' stands for the mixture of them. In the fourth/fifth  row, the Q-K/F-K attention maps for the second head of the second, fifth, seventh and eighth NBC blocks are given from the fourth to the seventh columns, respectively.}
  \label{fig:attn}
  \vspace{-0.2cm}
\end{figure*}

To study how NBC2 learns the spatial information to perform speech separation, in Fig. \ref{fig:attn}, we draw the attention scores of one head-tail overlapped speech pair produced by the proposed NBC2-small network. The first three rows and the first three columns of Fig. \ref{fig:attn} are the spectrogram of speech signal of the first speaker (`spk1') and the second speaker (`spk2'), and their mixture (`mix'), respectively. 
Let's denote the attention score of one head of one layer for all frequencies as ${S}_{f,q,k} \in [0,1]$, where $f\in\{0,...,F-1\}$, $q\in\{1,...,T\}$ and $k\in\{1,...,T\}$ denote the indices of frequency, query and key, respectively, with $\sum_k {S}_{f,q,k}=1$ according to the softmax function along $k$. 
In the fourth row, we draw the Q-K attention map of the second head of the second, fifth, seventh and eighth NBC blocks, from the fourth to the seventh columns, respectively. The Q-K attention maps draw the  attention scores averaged over frequencies, i.e. $\tilde{S}_{q,k} = \frac{1}{F}\sum_f {S}_{f,q,k}$,  and reflect the attentions between frames. 
In the fifth row,, we draw the F-K attention maps of the same heads as the Q-K attention maps. The F-K attention maps draw the attention scores averaged over queries, i.e. $\bar{S}_{f,k} = \frac{1}{T}\sum_q {S}_{f,q,k}$, and reflects the contribution of each TF bin (to the TF bins at the same frequency).
Note that we clipped the Q-K attention scores to have a maximum value of 0.03 to make the attention patterns more clearly visible. 
%The Q-K attention images show the attention patterns between different time steps, and the F-K attention images show the contribution of each TF bin among the TF bins of the same frequency.
% For convenience, we call the frequency-averaged attention image, i.e. the image of $\mathbf{S}_{q,k}$, as Q-K image and call the query-averaged attention image, i.e. the image of $\mathbf{S}_{f,k}$, as F-K image.

From Fig. \ref{fig:attn}, we observe two interesting points about how the proposed NBC2 network works:
\begin{itemize}[leftmargin=*]
\item \emph{Speaker Clustering.} 
At the lower layers, the first Q-K map (for the second NBC block) shows that all the frames of two speakers attend to each other, and the two speakers are not well separated. While, at the higher layers, the second and fourth Q-K maps (for the fifth and eighth NBC blocks, respectively) show very strong speaker clustering patterns, as each head attends to only one of the two speakers. This can also be verified by the second and fourth F-K maps. This type of single-speaker head commonly presents at the higher layers of the network. %The second and the third Q-K and F-K images show very strong clustering patterns focusing on the second and the first speakers respectively, which is common in other examples in high layers of the network.
Besides, at the higher layers, as shown in the third Q-K attention map (for the seventh NBC block), there also exists a few heads that both the two speakers are present in one head, but the frames of one speaker mainly attend to the frames of the same speaker.
Overall, the proposed network is indeed performing speaker clustering based on the frame-wise spatial vectors, and the speaker clustering is gradually completed from the lower to the higher layers. 
%The heads in low layers of the network like the second head in the second layer shown in the first Q-K and F-K images don't conduct obvious speaker clustering as high layers do, instead they seem to focus more on the TF bins with clear spatial information.
%For example, in the first F-K image, from the starting frames of each phoneme to its reverberant part, the attention score decays fast relative to the energy decaying speed in the spectrograms.
%The starting frames generally have clear spatial information not affected by reverberation effect.
%Thus, we can say that low layers in the proposed NBC2 network focus more on the TF bins with clear spatial information, while in high layers, the network conducts speaker clustering to perform speech separation.
\item \emph{Reverberations Effect.} 
For this example utterance, RT60 is 0.8 s and thus reverberation is heavy. 
At the lower layers, as shown in the first Q-K map, the vertical attention lines mainly locate around the onsets of speech components. We can also observe from the first F-K map that only the onsets of speech components are notably attended. According to the precedence effect \cite{litovsky1999precedence}, the onsets of speech components are mainly composed of the direct-path propagation of sound as  reflections have not arrived yet. In other words, the network starts to learn knowledge from the TF bins that are less contaminated by reverberation. At the higher layers, as shown in the second Q-K map, the vertical attention lines become (visibly) wider relative to the ones in the first Q-K map. The second F-K  map also shows that more reverberant TF bins are attended. When the layer get further higher, 
%The attention pattern in first Q-K image is thin vertical lines compared with other Q-K images, and the first F-K image doesn't have long and strong reverberant tails as the mixture spectrogram, both of which indicates that this head, i.e. the second head of the second layer, doesn't give too much attention to the reverberation in the speech mixture.
% The attention pattern in first Q-K image (the second head of the second layer) is thin vertical lines compared with other Q-K images, which indicates that this head doesn't give too much attention to the reverberation in the speech mixture.
% The corresponding F-K image, i.e. the first one, also verifies that. The attention pattern in first F-K image is a modelling of the spectral pattern presented in the mixture spectrogram, but it doesn't have long and strong reverberant tails as in the mixture spectrogram.
%The reverberation parts get larger attention weights with the increase of layer index, as the vertical lines become wider in second Q-K image.
in the third and fourth Q-K maps, the attentions appear to be some slashes corresponding to the reverberation effect. The third and fourth F-K maps also show that even more reverberant TF bins are attended relative to the second F-K map. 
The slashes are well temporally structured even though the proposed network does not use any type of positional embedding. We would like to note that these kind of slashes generally become longer when RT60 becomes larger, and are less prominent in low reverberation examples, thus the slashes can be considered as a proper modelling of reverberation.
Overall, we can conclude that reverberation is modelled gradually from the lower to the higher layers.
%This conclusion can be testified by the corresponding F-K images, in which the attention scores of the reverberant tails become relatively larger as the increase of layer index.
\end{itemize}

\section{Conclusions}
In this paper, we propose a multichannel narrow-band speech separation network, i.e. the revised narrow-band conformer (NBC2). The frequency permutation problem is solved by using the full-band PIT training strategy, which binds the same output position of all frequencies.
In NBC2, the self-attention mechanism clusters the frame-wise spatial vectors by measuring their similarities. The enhanced convolutional layers are expected to compute speech statistics and model reverberation. The attention maps demonstrate that the reverberation effect is indeed gradually modelled by the proposed network. 
Besides, group batch normalization (GBN) is proposed to account for the high correlation of hidden units across frequencies, and it largely improves the speech separation performance relative to other normalization methods. 
%GBN is proposed for the case where the whole network or partial of it can see only partial
%information of one group of tightly related training samples, as the case in the proposed method.
%We conducted experiments to compare performance of GBN and its rivals layer normalization, batch normalization and group normalization.
%Our experiments show that in the narrow-band separation task, GBN performs much better over the three normalization methods (more than 3 dB SDR and 0.5 WB-PESQ improvement).
Experiments show that the proposed method works well for various speech overlap ways and microphone array settings, and outperforms other state-of-the-art methods by a large margin. In addition, the proposed narrow-band method is spectrum-agnostic, and can be well-trained using only one hour of four speakers' data. The excellent performance of the proposed NBC2 network verifies that the narrow-band spatial information are highly discriminative for speech separation, and the proposed network is effective to fully leverage these information. 

Although the proposed network achieves satisfying speech separation performance on the 
simulated two-speaker datasets in this work, the performance may degrade for more challenging scenarios, such as with more speakers, moving speakers and/or background noise, or on real-recorded data. The proposed network only exploits the narrow-band spatial information, which can be integrated with  full-band spectral/spatial information to further improve the performance, as future work. The fusion of full-band and sub-band/narrow-band have already been studied in \cite{hao_FullsubnetFullBandSubBand_2021,tesch2022insights,yang2022mcnet}, but for LSTM networks. In addition, this work only considers short utterances, and extending the proposed network for continuous speech separation will be done in the future.

\bibliographystyle{IEEEtran}

\bibliography{mybib}


%\end{IEEEbiography}


%\newpage



%\vfill


\end{document}